\documentclass{aa}  
\usepackage{graphicx}
\usepackage{txfonts}
\begin{document}
   \title{Turbulent Driving Scales in Molecular Clouds}

   \author{C. M. Brunt,
          \inst{1}
	   M. H. Heyer,
	  \inst{2}
          \and
	  M.-M. Mac Low
	  \inst{3}
          }

   \offprints{C. M. Brunt}

   \institute{Astrophysics Group, School of Physics, University of Exeter, Stocker Road, Exeter, EX4
   4QL, UK\\
              \email{brunt@astro.ex.ac.uk}
         \and
             Department of Astronomy, University of Massachusetts at Amherst, 710 North Pleasant
	     Street, Amherst, MA~01003, USA\\
             \email{heyer@astro.umass.edu}
	 \and
	     Department of Astrophysics, American Museum of Natural History,
	     79$^{th}$ Street and Central Park West, New York, NY~10024--5192, USA\\
	     \email{mordecai@amnh.org}
             }

   \date{Received ; accepted }

 
  \abstract
   { Supersonic turbulence in molecular clouds is a dominant agent that strongly affects the clouds' 
     evolution and star formation activity. Turbulence may be initiated and maintained by a number of
     processes, acting at a wide range of physical scales. By examining the dynamical state of molecular
    clouds, it is possible to assess the primary candidates for how the turbulent energy is injected. }
   { The aim of this paper is to constrain the scales at which turbulence is driven in the molecular
       interstellar medium, by comparing simulated molecular spectral line observations of numerical 
       magnetohydrodynamic (MHD) models and molecular spectral line observations of real molecular clouds.} 
   { We use principal component analysis, applied to both models and observational data, to extract a
       quantitative measure of the driving scale of turbulence. } 
   { We find that only models driven at large scales (comparable to, or exceeding, the size
       of the cloud) are consistent with observations. This result applies also to clouds with little or
       no internal star formation activity.} 
    { Astrophysical processes acting on large scales, including supernova-driven turbulence,
      magnetorotational instability, or spiral shock forcing, are viable candidates for the generation
      and maintenance of molecular cloud turbulence. Small scale driving by sources internal to molecular
      clouds, such as outflows, can be important on small scales, but cannot replicate the observed 
      large-scale velocity fluctuations in the molecular interstellar medium.}  


   \maketitle
%

\section{Introduction}

Turbulence is an important agent that controls the evolution (and perhaps formation) of
molecular clouds and the subsequent production of stars. As such, it has attracted
significant attention from theorists, especially since the advent of numerical supercomputer
simulations. Of particular interest is the source(s) of energy injection that
create and sustain turbulence in molecular clouds. A number of different mechanisms
have been proposed, including supernovae, H{\sc II} regions, outflows, spiral arms,
magneto-rotational instability in galactic disks (Mac Low \& Klessen 2004; Miesch \& Bally 1994). 
These mechanisms may be distinguished by the effective spatial scale at which 
they preferentially operate, and clues to the nature of the energy injection mechanism(s) 
may be extracted from spectral line imaging observations of molecular clouds.  

A number of methods for studying resolved velocity fields in molecular clouds
have been developed and applied. These include projected velocity 
(line centroid) analysis (e.g. Scalo 1984; Miesch \& Bally 1994;
Ossenkopf \& Mac Low 2002; Brunt \& Mac Low 2004), the spectral correlation function (SCF; Rosolowsky 
{\it et al}{} 1999), velocity channel analysis (VCA; Lazarian \& Pogosyan 2001, 2004), and
principal component analysis (PCA; Heyer \& Schloerb 1997). 
To date, these methods have been used to estimate the power law indices of the velocity
structure function/power spectrum in molecular clouds from observed data cubes of
molecular line emission  (e.g. Brunt \& Heyer 2002b; Heyer \& Brunt 2004).

Application of PCA to Outer Galaxy molecular clouds (Brunt 2003a -- Paper I hereafter) 
revealed that, in comparison to simple models, the observational record favoured 
large scale driving of turbulence in the molecular clouds. In their study of the Polaris
molecular cloud, Ossenkopf \& Mac Low (2002) also found that large scale driving of turbulence
provided a better explanation of the cloud's velocity structure.

In this paper, we construct simulated observations of molecular clouds, derived from 
computational simulations of interstellar turbulence. The models include magnetic fields 
and self-gravity and are driven (randomly forced) on a range of spatial scales.
We employ PCA to quantitatively investigate the observational signatures of different driving scales. 
Our numerical measurements are compared to previous PCA results obtained from the simple cloud models
of Paper I and to the same measurements made on real molecular clouds. The layout of the paper is
as follows. In Section~2, we briefly summarize the PCA method and review the relevant findings of
Paper I. Section~3 introduces the numerical models and summarizes the simulated observations of
these. In Section~4, we present our results, compare these to corresponding observations, 
and discuss the implications for the generation of turbulence in molecular clouds. Our 
conclusions are given in Section~5.


\section{Principal Component Analysis}

Principal component analysis can be used to decompose three dimensional spectral line imaging 
observations onto orthogonal spectroscopic eigenvectors along which
ordered sources of variance in the data are maximized (Heyer \& Schloerb 1997). 
Projection of the data onto the eigenvectors produces a sequence of diagnostic eigenimages.
We refer below to each coupled eigenvector-eigenimage pair as a principal component (PC), distinguished by
its order $m$~=~1,2,..,N, where N is the number of spectroscopic channels of the data set.  The amount of 
variance in the data accounted for by the PCs is a decreasing function of $m$. The characteristic sizes of
eigenimage structures are measured as the spatial scale at which their autocorrelation function (ACF) falls 
to 1/$e$ of the zero-lag value (Brunt \& Heyer (2002a). At order $m$ we denote the characteristic 
spatial scale of the eigenimage as $l_{m}$.

In the literature there are numerous examples of eigenimages obtained from simple molecular cloud models 
and from observations of real molecular clouds (e.g. Heyer \& Schloerb 1997; Brunt 1999; Brunt 2002b). 
Real molecular cloud eigenimage sequences display chaotic structures that are only replicated by models
that contain chaotic (turbulent) velocity fluctuations on all scales. A quantitative statement on this
was given in Paper I, as summarized below.

The analysis of Paper I considered fractional Brownian motion velocity
fields with correlated velocity fluctuations up to a maximum size scale defined by
the turnover wavenumber, $k_{cut}$, in the velocity power spectra; $k_{cut}$ determines
the largest wavelength, $\lambda_{0}$, at which correlated velocity fluctuations are present.
For wavenumbers greater than $k_{cut}$, the power spectrum
was a power law, while for wavenumbers less than $k_{cut}$ the power spectrum was flat.
For the simple models of Paper I, $\lambda_{0}$ is used as a surrogate for the driving scale,
$\lambda_{D}$. The model velocity fields of Paper I were then embedded in a ``cloud'' -- this
was simply a Gaussian density distribution parameterized by the spatial FWHM, L$_{c}$.
The combined density and velocity fields were then transferred to the observational
axes via a density-weighted projection of the line-of-sight velocity field. 

Upon applying PCA, it was found that the ratio of characteristic spatial scales, $l_{2}$/$l_{1}$,
derived from the first two eigenimages, was sensitive to variations in $\lambda_{0}$/L$_{c}$.
In detail : $l_{2}$/$l_{1}$ was tightly correlated with $\lambda_{0}$/L$_{c}$ for $\lambda_{0}$/L$_{c}$~$<$~1.
For models with  $\lambda_{0}$/L$_{c}$~$>$~1, no correlation of $l_{2}$/$l_{1}$ with $\lambda_{0}$/L$_{c}$ 
was observed, but all models with $\lambda_{0}$/L$_{c}$~$>$~1 could be readily distinguished from models with
$\lambda_{0}$/L$_{c}$~$<$~1. 

When these models were compared to spectral line observations of real molecular clouds, it was
found that only models which included large scale velocity fluctuations could match the observational 
data. We now repeat the analysis of Paper I using more realistic molecular cloud models obtained via
numerical simulation of driven turbulence. Radiative transfer of $^{12}$CO and $^{13}$CO (J=1--0)
spectral lines was included in the construction of the ``observable'' models. Both of these
features are an advance over Paper I. 


\section{Numerical Data}

\subsection{Overview~{\label{sec:overview}}}

We use simulations of randomly driven hydrodynamical (HD) turbulence and magnetohydrodynamical (MHD) 
turbulence (Mac Low 1999), performed with the astrophysical MHD code 
ZEUS-3D\footnote{Available from the Laboratory for Computational Astrophysics at http://lca.ucsd.edu/portal/software/zeus-3d} 
(Clarke 1994), a 3D version of the code described by Stone \& Norman (1992a, b). Further details on the
numerical scheme are found in Mac Low (1999). To drive the turbulence, a fixed pattern of
Gaussian fluctuations is drawn from a field with power only in a narrow band of wavenumbers around 
some value $k_d$. The dimensionless wavenumber(s) $k_d$, at which the simulations are driven, 
counts the number of driving wavelengths $\lambda_d$ in the computational box.
This pattern is normalized to produce a set of perturbations that are added to the
velocity field, with the amplitude chosen to maintain constant kinetic energy input rate.
This offers a very simple approximation to driving by mechanisms that act on a particular scale.
In general, one must recognize the possibility of multi-scale energy injection from a variety of 
sources (Scalo 1987). However, for our purposes here, the numerical simulations provide a
conveniently parameterized sample of ``clouds'' with which to investigate the observational signatures 
of different driving scales. We also include models with self-gravity in which the turbulence is driven 
at small and large scales (Klessen, Heitsch, \& Mac Low 2000). In these models, turbulence is initiated 
in the fluid and allowed to reach steady state before self-gravity is turned on. We include snapshots of 
these models at a number of timesteps ($t/t_{ff}$~=~0,~1,~5/3) where $t_{ff}$ is the free-fall timescale
and $t$~=~0 refers to the point at which self-gravity is turned on.

A summary of the models is given in Table~\ref{table:1}. The models are scale free; we impose physical
units as follows : mean density n$_{H_{2}}$~=~139~cm$^{-3}$; linear size L~=~10~pc; 
sound speed c$_{s}$~=~0.265~km~s$^{-1}$ (T$_{k}$~=~17~K; isothermal) --
see Mac Low (1999), Klessen, Heitsch, \& Mac Low (2000). All simulations were performed on a 128$^{3}$ grid. 

\subsection{Simulated Observations~~{\label{sec:simobs}}}

To generate observed simulations directly comparable to real data, we apply
radiative transfer calculations to the numerically simulated velocity and density fields 
The physical fields are transferred onto the observational axes using a non-LTE
excitation calculation that accounts for local radiative trapping at each grid point,
followed by radiative transfer through the grid (see Brunt \& Heyer 2002a).  The intensities of the 
$^{13}$CO and $^{12}$CO spectral lines are computed at velocity resolution 0.05~km~s$^{-1}$.

The ``cloud size'' for the simulations is, nominally, the size of the computational
box. However, the simulated density fields (particularly for small scale driving) do not have sufficient 
(column) density contrast to enable a meaningful measurement of $l_{1}$
because the ACF of the first eigenimage does not fall to the $1/e$ point to which the
spatial scale measurements are referenced. This could be avoided by padding the fields before
ACF computation, but this is a poor choice as the fields are actually periodic.
In order to ensure a more meaningful ``cloud size'' for the models, we have defined a spherical
window of 100~pixels diameter within the computational box.
Within this window, the density field is taken as simulated, and we taper to zero density
quickly but smoothly outside this window. We take the ``cloud size'' as the diameter of the imposed
spherical window, denoted as $L_{c}$. 

The driving wavelength, $\lambda_{D}$ is $N_{pix}/(min)\lambda_{d}$ where $N_{pix}$~=~128
and $(min)\lambda_{d}$ is the smallest wavenumber within the driving range (i.e. 1, 3, or 7;
see Table~\ref{table:1}). This results in values of $\lambda_{D}$~=~128, 42.7 and
18.3, and values of the ``fractional driving scale'' $\lambda_{D}$/L$_{c}$~=~1.28, 0.427, and 0.183.
 

\section{Turbulent Driving Scales~{\label{sec:TDS}}}

\subsection{Results~{\label{sec:results}}}

\begin{figure}
\centering
\includegraphics[angle=0, width=9.5cm]{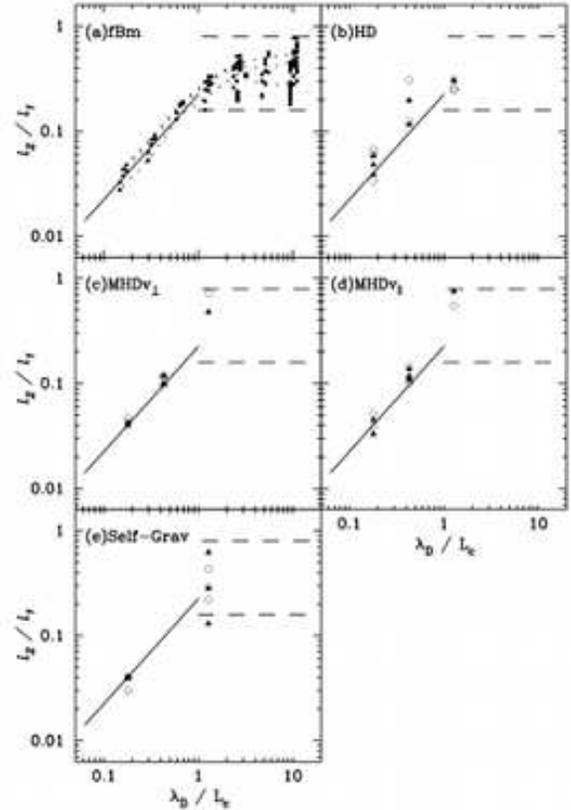}
   \caption{(a) Plot of the ratio of scales from the second and first eigenimages, $l_{2}/l_{1}$,
   versus the driving scale ratio $\lambda_{D}/L_{c}$ (dots) obtained from PCA of simple model
   clouds (Paper I). The solid line marks the trend of $l_{2}/l_{1}$ with $\lambda_{D}/L_{c}$ for
   $\lambda_{D}/L_{c} < 1$. The dashed lines mark the range of $l_{2}/l_{1}$ observed when
   $\lambda_{D}/L_{c} > 1$. (b) Plot of $l_{2}/l_{1}$ versus $\lambda_{D}/L_{c}$ for the HD
   simulations. (c) Plot of $l_{2}/l_{1}$ versus $\lambda_{D}/L_{c}$ for MHD ($v_{\perp}$). 
   (d) Plot of $l_{2}/l_{1}$ versus $\lambda_{D}/L_{c}$ for MHD ($v_{||}$). (e) Plot of 
   $l_{2}/l_{1}$ versus $\lambda_{D}/L_{c}$ for the self-gravitating models. In panels (b)-(e)
   $^{12}$CO and $^{13}$CO measurements are represented by open circles and triangles
   respectively. }
      \label{fig:lolc}
\end{figure}

PCA was applied to the simulated observations according to the procedure given in Brunt \& Heyer (2002a).
The characteristic spatial scales, $l_{1}$ and $l_{2}$, are derived from the first two eigenimages 
of each simulation and the ratios, $l_{2}/l_{1}$, are listed in Table~\ref{table:1}.
Figure~\ref{fig:lolc} shows these measurements plotted against $\lambda_{D}$/L$_{c}$ and compared to the
simple cloud results of Paper I. The $l_{2}/l_{1}$ measurements from the numerical models are in
good agreement with the simple fBm cloud results of Paper I. Figure~\ref{fig:lolc} verifies that $l_{2}/l_{1}$
provides a coarse measure of the turbulent driving scale. Note that for $\lambda_{D}$/L$_{c}$~$>$~1 
($l_{2}/l_{1}$~$>$~0.1--0.2) there is little or no sensitivity to the actual driving scale, and this
regime should be viewed simply as ``large scale driving''. According to the results of Paper I, the
variation of $l_{2}/l_{1}$ between $\sim$~0.2--0.8 occurs naturally, due to the unpredictability of the
projection of a large scale velocity gradient onto the line of sight. The results presented here show
that the magnetic model driven at large scales, ME21, has a larger $l_{2}/l_{1}$ than the hydrodynamic
models HC2, HE2. In light of the Paper I results, not too much should be read in to this result without
further study. Similarly, temporal variations in $l_{2}/l_{1}$ for the D1H model should not be over-interpreted.

\begin{table*}
\begin{minipage}[t]{\textwidth}
\caption{Numerical Models: Parameters and PCA Measurements}             
\label{table:1}      
\centering   
\renewcommand{\footnoterule}{}  
\begin{tabular}{lcccccccc}
\hline \hline
Model & $k_{d}$\footnote{Driving wavenumber} & M\footnote{rms Mach number} &
$v_{A}/c_{s}$\footnote{Ratio of Alfv{\'e}n speed to sound speed} & $\lambda_{D}$/L$_{c}$ & $l_{2}/l_{1}$ ($^{13}$CO) & $l_{2}/l_{1}$ ($^{12}$CO)\\
\hline
HA8  &                  7--8 &  1.9  &  0 & 0.18  & 0.07 $\pm$ 0.01 & 0.05 $\pm$ 0.01 \\
HC8  &                  7--8 &  4.1  &  0 & 0.18  & 0.06 $\pm$ 0.01 & 0.06 $\pm$ 0.01 \\
HE8  &                  7--8 &  8.7  &  0 & 0.18  & 0.03 $\pm$ 0.01 & 0.04 $\pm$ 0.01 \\  
HC4  &                  3--4 &  5.3  &  0 & 0.43  & 0.12 $\pm$ 0.02 & 0.12 $\pm$ 0.02 \\
HE4  &                  3--4 & 12.0  &  0 & 0.43  & 0.31 $\pm$ 0.04 & 0.20 $\pm$ 0.02 \\
HC2  &                  1--2 &  7.4  &  0 & 1.28  & 0.25 $\pm$ 0.03 & 0.30 $\pm$ 0.03 \\
HE2  &                  1--2 & 15.0  &  0 & 1.28  & 0.26 $\pm$ 0.03 & 0.31 $\pm$ 0.03 \\
MC81 $v_{||}$       &   7--8 &  3.5  &  1 & 0.18  & 0.05 $\pm$ 0.01 & 0.05 $\pm$ 0.01 \\
MC81 $v_{\perp}$    &   7--8 &  3.5  &  1 & 0.18  & 0.05 $\pm$ 0.01 & 0.04 $\pm$ 0.01 \\
MC85 $v_{||}$       &   7--8 &  3.4  &  5 & 0.18  & 0.04 $\pm$ 0.01 & 0.03 $\pm$ 0.01 \\
MC85 $v_{\perp}$    &   7--8 &  3.4  &  5 & 0.18  & 0.04 $\pm$ 0.01 & 0.04 $\pm$ 0.01 \\ 
MC41 $v_{||}$       &   3--4 &  4.7  &  1 & 0.43  & 0.10 $\pm$ 0.02 & 0.14 $\pm$ 0.02 \\
MC41 $v_{\perp}$    &   3--4 &  4.7  &  1 & 0.43  & 0.10 $\pm$ 0.02 & 0.10 $\pm$ 0.02 \\
MC45 $v_{||}$       &   3--4 &  4.8  &  5 & 0.43  & 0.14 $\pm$ 0.02 & 0.11 $\pm$ 0.02 \\
MC45 $v_{\perp}$    &   3--4 &  4.8  &  5 & 0.43  & 0.12 $\pm$ 0.02 & 0.10 $\pm$ 0.02 \\ 
MC4X $v_{||}$       &   3--4 &  5.3  & 10 & 0.43  & 0.12 $\pm$ 0.02 & 0.12 $\pm$ 0.02 \\
MC4X $v_{\perp}$    &   3--4 &  5.3  & 10 & 0.43  & 0.11 $\pm$ 0.02 & 0.12 $\pm$ 0.02 \\ 
ME21 $v_{||}$       &   1--2 & 14.0  &  1 & 1.28  & 0.55 $\pm$ 0.06 & 0.75 $\pm$ 0.08 \\
ME21 $v_{\perp}$    &   1--2 & 14.0  &  1 & 1.28  & 0.72 $\pm$ 0.09 & 0.48 $\pm$ 0.05 \\ 
D1H($t/t_{ff}$=0)   &   1--2 & 10.0  &  0 & 1.28  & 0.62 $\pm$ 0.03 & 0.43 $\pm$ 0.04 \\
D1H($t/t_{ff}$=1)   &   1--2 & 10.0  &  0 & 1.28  & 0.13 $\pm$ 0.03 & 0.29 $\pm$ 0.04 \\
D1H($t/t_{ff}$=5/3) &   1--2 & 10.0  &  0 & 1.28  & 0.28 $\pm$ 0.06 & 0.22 $\pm$ 0.02 \\
D3H($t/t_{ff}$=0)   &   7--8 & 10.0  &  0 & 0.18  & 0.04 $\pm$ 0.01 & 0.03 $\pm$ 0.01 \\
D3H($t/t_{ff}$=1)   &   7--8 & 10.0  &  0 & 0.18  & 0.04 $\pm$ 0.01 & 0.04 $\pm$ 0.01 \\
D3H($t/t_{ff}$=5/3) &   7--8 & 10.0  &  0 & 0.18  & 0.04 $\pm$ 0.01 & 0.04 $\pm$ 0.01 \\
\hline
\end{tabular}
\end{minipage}
\end{table*}

\begin{figure}
\centering
\includegraphics[width=9cm]{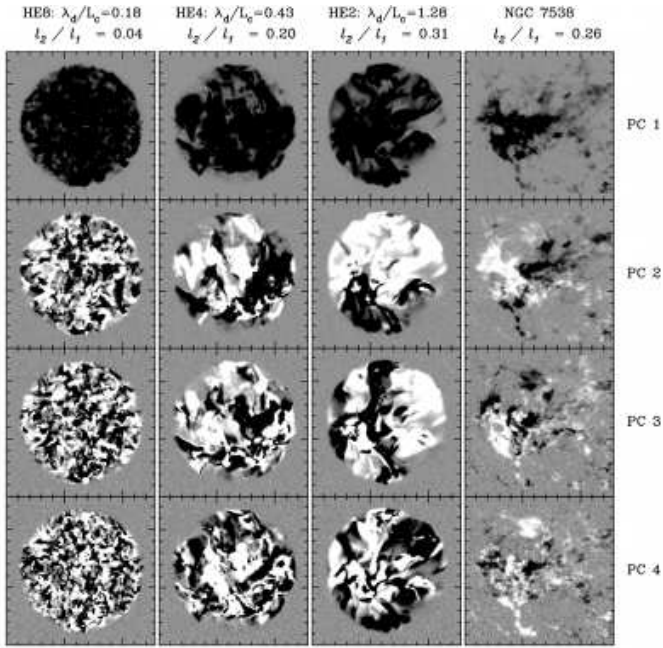}
   \caption{Example $^{12}$CO eigenimage sequences for the first four principal
   components obtained from the hydrodynamical data for $\lambda_{D}$/L$_{c}$=0.18, 0.43, and 1.28 and from 
   observations of the NGC~7538 molecular cloud. The eigenimages of molecular 
   clouds are more similar to the numerical cloud models with turbulence driven 
   at scales comparable to the cloud size.
	   }
      \label{fig:pcseq}
\end{figure}

A visual example of the data presented in Table~\ref{table:1} is given in
Figure~\ref{fig:pcseq}, where we display the first 4 eigenimages obtained from
$^{12}$CO simulated observations of HE2, HE4, and HE8 (c.f. Figure~3 of Paper I).
Figure~\ref{fig:pcseq} also includes the first 4 eigenimages obtained from 
$^{12}$CO observations of the NGC~7538 molecular cloud, for which
$l_{2}/l_{1}$~=~0.26~$\pm$~0.09. Figure~\ref{fig:pcseq} demonstrates 
that for turbulence driven on small scales, the higher order ($m>1$) eigenimage structures are
confined to small scales relative to the overall cloud size.  Cloud models with large 
scale driving of turbulence generate large second eigenimage structures with respect to
the overall cloud size, typically displaying a positive-negative ``dipole'' structure. 
The ratio $l_{2}/l_{1}$ is a simple quantitative measure of this trend.

\begin{figure}
\centering
\includegraphics[width=9cm]{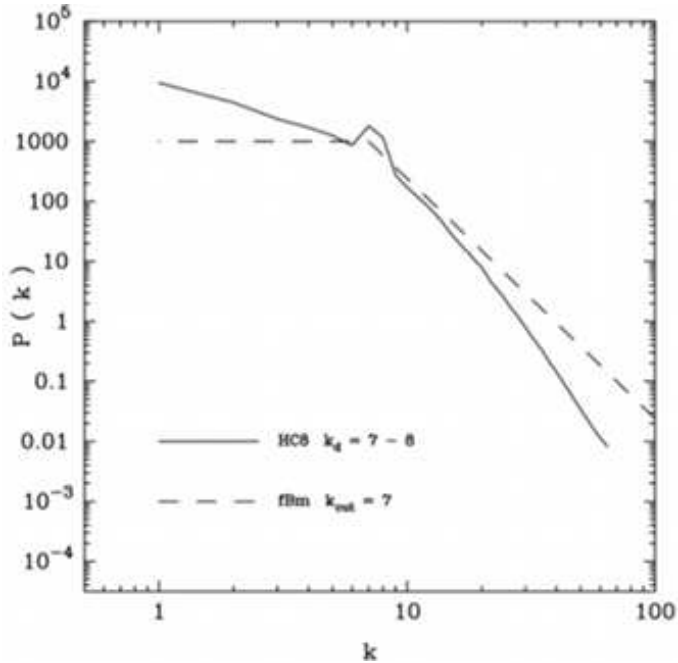}
   \caption{A comparison of velocity power spectra (power $P$ versus wavenmuber $k$) obtained from
            a numerically simulated cloud (HC8, with $k_{d}$~=~7--8) and an fBm field
            ($k_{cut}$~=~7) of the type used in Paper I to represent turbulent driving at
            $k$~$\approx$~7. HC8 has more power at low wavenumbers relative to the fBm field.
            (The vertical scale in this plot is arbitrary.)
           }
      \label{fig:powerspec}
\end{figure}

It is evident from Figure~\ref{fig:lolc} that there is a small trend for the recovered
$l_{2}/l_{1}$ to be very slightly larger than the results found for the fBm fields
of Paper I (this is most evident in panels (b) and (d) of Figure~\ref{fig:lolc}). Inspection
of the power spectra of the model velocity fields reveals the likely origin of this
effect. In Paper I, the fBm velocity fields were designed to have a power law spectrum
at wavenumbers greater than a cut-off wavenumber, $k_{cut}$; above this wavenumber, the
power was flat (independent of $k$). The numerically simulated velocity fields, on the
other hand, have excess power relative to the fBm models at low wavenumbers. A representative
example of this is demonstrated in Figure~\ref{fig:powerspec} where the model HC8, driven at
$k_{d}$~=~7--8 is compared to an ``equivalent'' fBm model with $k_{cut}$~=~7. While an
obvious turnover in spectral power is clearly evident at $k$~$<$~$k_{d}$ for HC8, it is
not as sharp as the corresponding fBm field that used as a surrogate in Paper I. 
Figure~\ref{fig:lolc} demonstrates, however, that the driving scale is still recoverable
using PCA.

\begin{figure}
\centering
\includegraphics[width=9cm]{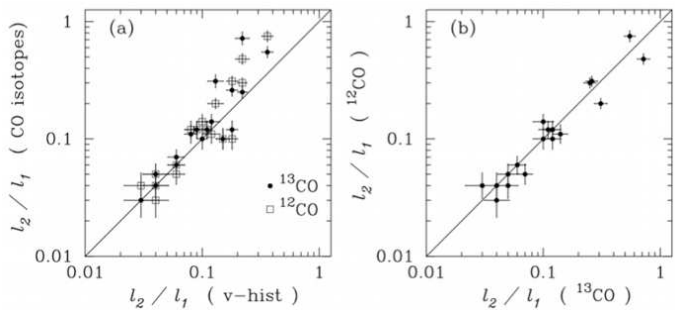}
   \caption{(a) Comparison of $l_{2}$/$l_{1}$ derived from simulated CO observations and observations
            using the $v$-hist method where opacity and excitation effects are not included.
            (b) Comparison of $l_{2}$/$l_{1}$ derived from simulated $^{13}$CO and $^{12}$CO observations. For each p
lot, the solid line denotes equivalent values along the ordinate and absissca axes.
           }
      \label{fig:isotope}
\end{figure}

Another important consideration is the effect of radiative transfer of the spectral lines. 
To investigate this, density-weighted velocity histograms (e.g. Falgarone et al. 1994) were 
also constructed as an approximation to a perfectly-excited optically thin spectral line
observation (referred to below as ``$v$-hist'' models). The $v$-hist models provide a baseline for 
investigating the effect of saturation on the analysis. As the $v$-hist models include no 
saturation effects, we used these to examine any biases arising from the use of 
$^{13}$CO and $^{12}$CO where opacity and excitation effects are present.
Figure~\ref{fig:isotope}(a) compares $l_{2}/l_{1}$ derived from $^{13}$CO and $^{12}$CO with
$l_{2}/l_{1}$ derived from the $v$-hist method. At small $l_{2}/l_{1}$ (i.e. small $\lambda_{D}$/L$_{c}$)
there are no systematic effects arising from opacity in the spectral lines. However, at higher
$l_{2}/l_{1}$ the CO emission overestimates $l_{2}/l_{1}$ relative to $v$-hist. 
This effect starts to become evident at $l_{2}/l_{1}$~$\approx$~0.1--0.2, which, as shown in
Figure~\ref{fig:lolc}, is the point at which $\lambda_{D}$/L$_{c}$~$\approx$~1 (i.e. the turbulence
is driven at the scale of the cloud). For $l_{2}/l_{1}$ (or $\lambda_{D}$/L$_{c}$) greater than 
this transition point, $l_{2}/l_{1}$ not surprisingly loses any sensitivity to the actual driving scale.
We conclude that there are no serious problems arising from opacity effects, and simply note that
values of $l_{2}/l_{1}$ greater than $\sim$0.1--0.2 are indicative of large scale driving
of turbulence. Interestingly, values of $l_{2}/l_{1}$ derived from real molecular clouds can
significantly exceed 0.2 (this is not typically seen in $v$-hist models), and we identify the source
of this as opacity effects in clouds driven at large scales. In Figure~\ref{fig:isotope}(b) we
compare $l_{2}/l_{1}$ derived from $^{13}$CO and $^{12}$CO. While noting the difference between
the CO observations and $v$-hist observations, there is clearly no systematic difference found
between $^{13}$CO and $^{12}$CO. This is in accord with previous investigations of PCA for
other applications (Brunt 2003b). 
Finally, we note that the inclusion of self-gravity does not
have any effect on the observed $l_{2}/l_{1}$, as can be seen in Figure~\ref{fig:lolc}.

\subsection{Discussion~{\label{sec:discussion}}}

\begin{figure} 
\centering
\includegraphics[width=9cm]{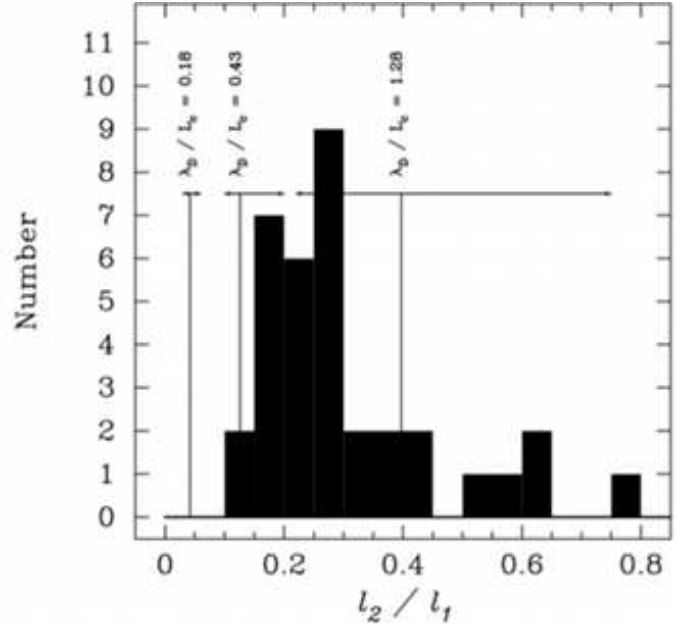}
   \caption{Histogram of $l_{2}$/$l_{1}$ obtained from $^{12}$CO observations of real molecular clouds. The vertical
           lines mark the mean $l_{2}$/$l_{1}$ derived from the model observations ($^{12}$CO) and the horizontal
           arrows extend over the range of measured $l_{2}$/$l_{1}$.
           }
      \label{fig:lhist}
\end{figure}

Both analytical and computational descriptions of turbulence are necessarily constrained by observations of 
interstellar clouds. A qualitative inspection of Figure~\ref{fig:pcseq} shows that the eigenimages derived from 
clouds models with large $\lambda_{D}$/L$_{c}$ are more consistent with the observations of NGC~7538.
More generally, the measured values of $l_{2}$/$l_{1}$ from real molecular clouds are typically $\gtrsim$~0.2.
In Figure~\ref{fig:lhist} we plot the histogram of $l_{2}$/$l_{1}$ measured in the sample of 
clouds from Paper I, to which we have added additional measurements from the clouds 
analyzed in Heyer \& Brunt (2004). In the combined sample there are 35 clouds in total.
Using Figure~\ref{fig:lolc} as a guide to the relationship between 
$<l_{2}$/$l_{1}>$ and $\lambda_{D}$/L$_{c}$, these values imply that the molecular clouds 
are dominated by turbulence driven on large scales compared to the cloud sizes.
This may be simply a result of the driving scale itself determining the size of molecular clouds
(Ballesteros-Paredes \& Mac Low 2002; Paper I).

In our experiment, we have considered the simplified case where a single ``driving scale'' is in operation.
Within this limitation we identify large scale driving as the dominant scenario. In reality, turbulence
can in principle be driven on multiple scales by a number of mechanisms (Scalo 1987). The origin of large-scale energy 
injection is discussed by Mac Low \& Klessen (2004), who concluded that field supernovae were the dominant 
mechanism in regions where they occur, while magneto-rotational instability (Kim, Ostriker, \& Stone 2003;
Tamburro {\it et al}{} 2009) may provide a background level. In addition to these, other possible mechanisms include forcing by 
shocks in spiral arm potentials; Dobbs \& Bonnell (2007) demonstrate that the scale-dependent 
velocity dispersion in molecular clouds can be replicated by simulated clouds in a galactic disk
with a fixed spiral arm pattern. Most of these processes likely require that the molecular cloud
turbulence is inherited from still larger scale motions in the atomic ISM (Elmegreen 1993, 
Ballesteros-Paredes {\it et al} 1999; Brunt 2003a). In this scenario, the ``driving'' of 
molecular cloud turbulence could simply be due to the continuous downward cascade of 
turbulent energy, that not only injects the turbulence but is also responsible for 
the (potentially rapid) molecular cloud formation in the first place 
(Bergin {\it et al} 2004; Glover \& Mac Low 2007). 
The presence of large scale turbulence in molecular clouds would be a natural, inevitable 
consequence of their formation, and their subsequent evolution can be significantly 
affected by dynamical events occurring in the larger scale ISM.

Energy injection on (initially) small scales by the spatio-temporally intermittent
development of outflows, stellar winds and H{\sc ii} regions within the cloud may not be well 
modeled by random forcing methods used in these simulations. These point-like injections of 
energy can expand their spheres of influence over time and may ultimately contribute to large scale 
turbulent motions. However, on large scales, these processes are disfavoured on energetic 
grounds (Mac Low \& Klessen 2004). While there is evidence that energy injection by outflows 
can be important over limited scales (e.g. Bally, Devine, \& Alten 1996; Knee \& Sandell 2000) it is 
unlikely that outflow-driven turbulence can explain the origin of molecular cloud
turbulence as a whole (Walawender, Bally, \& Reipurth 2005; Banerjee, Klessen, \& Fendt 2007). 
This is demonstrated by recent simulations of outflow-driven turbulence which reveal that
energy injection by outflows is not capable of creating turbulence at scales comparable to the cloud size.
Models of interacting outflows generated either randomly (Carroll {\it et al}{} 2008), or
self-consistently (Nakamura \& Li 2007) show that turbulence is only injected with an
effective driving scale of about 1/5 to 1/10 the size of the cloud ($\lambda_{D}/L_{c}$~$\approx$~0.1--0.2)
which is incompatible with our results as summarized in Figure~\ref{fig:lolc} and Figure~\ref{fig:lhist}.
The observable ratio $l_{2}$/$l_{1}$ is expected to lie in the range 0.02--0.05 when
$\lambda_{D}/L_{c}$~$\approx$~0.1--0.2, according to our modelling results. Additionally, the
cloud modelled by Nakamura \& Li (2007) is only 1.5~pc in size, and it is unclear whether the
effective driving scale would increase (for the same outflow parameterization) if a larger 
cloud was modelled. If the {\it fractional} driving scale of 0.1--0.2 is interpreted as a 
{\it physical} driving scale of 0.15--0.3~pc, then outflow-driven turbulence would be even less
effective in globally exciting turbulence in larger clouds. On the other hand, in larger clouds,
more massive and energetic outflows may be expected to be present, but it is not currently
clear how (or if) the effective fractional driving scale would increase.

An observational estimate of the effective driving scale of turbulence by outflows was found
by Swift \& Welch (2008). They inferred an energy injection scale of 0.05~pc for L1551, which is a small
fraction of the the overall cloud diameter of around 1.8~pc. Using this estimate, they found a rough
balance between the energy injection rate (from the outflows) and the turbulent dissipation rate,
with a characteristic injection/decay timescale of $\sim$~0.1~Myr, which is substantially less than the
inferred cloud age of $\sim$~4--6~Myr. We note here that some caution is required in interpreting 
the appearance of injection/decay balance for the outflow-driven turbulence.
The dissipation time scale of turbulence is proportional to the driving scale 
(Mac Low 1999). Swift \& Welch (2008), in calculating their dissipation rate, used a 
driving scale of 0.05~pc, and therefore their result shows primarily
that the energy injection through outflows is quickly dissipated on short time scales over short
length scales. If the L1551 cloud is, or has been, subject to large scale ($\gtrsim$~1.8~pc) 
driving of turbulence, then the large scale turbulence is controlled by a much longer dissipation
timescale of (1.8/0.05)~$\times$~0.1~Myr~$\approx$~3.6~Myr, which is more in line with the cloud 
age. The small-scale driving of outflows would then occur within the longer time evolution of the cloud, 
set by the longer dissipation time scales of the initial turbulence, injected on large scales.

\begin{figure}
\centering
\includegraphics[width=9cm]{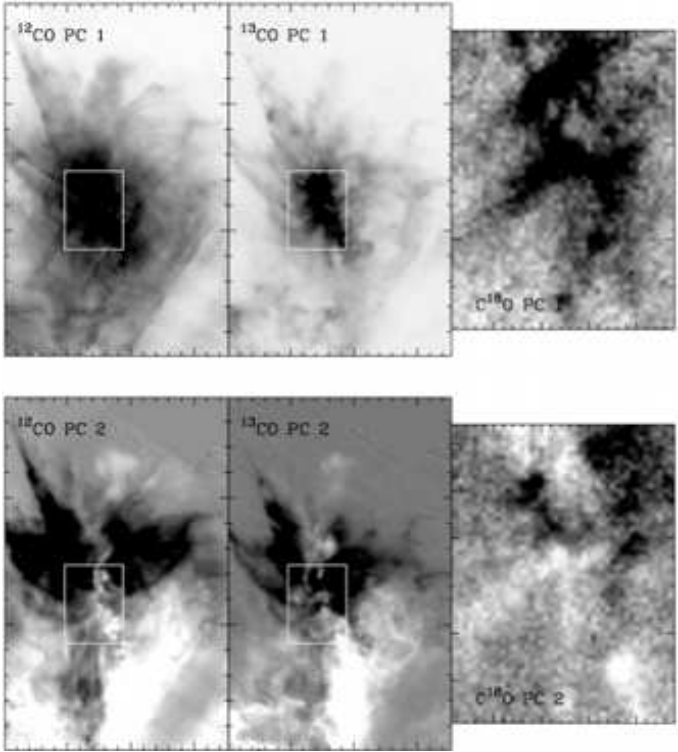}
   \caption{ First and second eigenmages obtained from principal component analysis of the NGC~1333 molecular cloud. The C$^{18}$O data and analysis are confined to the central core region, delineated by the rectangular box on the $^{12}$CO and $^{13}$CO images.}
      \label{fig:eigim1333}
\end{figure}

The effect of multiple outflows within a small region of space may be seen 
in the outflow-rich NGC~1333 molecular cloud. Here, Quillen {\it et al} (2005) 
describe as many 22 cavities within a 1~pc$^{3}$ volume, 
possibly excavated by outflow activity, in the $^{13}$CO (J=1--0) map of Ridge {\it et al} (2003). 
The cavities have typical diameters of 0.1--0.2~pc, indicative again of a small effective
driving scale, as the shells surrounding the cavities would presumably collide and merge at 
larger scales. However, it is not yet clear that outflows are the driving source for these
cavities, as many do not have obvious stellar sources inside them -- see Quillen {\it et al} (2005)
for further discussion. 

To investigate outflow-driven turbulence from an observational perspective, we
applied the PCA method to CO observations of the NGC~1333 molecular cloud. 
We used the J=1--0 spectral lines of $^{12}$CO and $^{13}$CO
observed at FCRAO as part of the COMPLETE project (Ridge {\it et al} 2006), as well as
FCRAO C$^{18}$O J=1--0 spectral line data towards the central core region of NGC~1333.
In Figure~\ref{fig:eigim1333} we show the first two eigenimages obtained from the analysis for
each spectral line. For $^{12}$CO and $^{13}$CO, we find ``dipole'' second eigenimage
structure characteristic of large scale turbulence, and measure $l_{2}$/$l_{1}$ values
of 0.59 ($^{12}$CO) and 0.63 ($^{13}$CO). The overall cloud size is estimated from the
$^{12}$CO $l_{1}$ measurement to be 3.27~pc, assuming a distance of 318~pc. 
These measurements show that turbulence is (or has been)
driven on large scales in NGC~1333, and is unlikely to have originated from the outflows, which
are confined to the central core region, marked by the small box in Figure~\ref{fig:eigim1333}.  
Analysis of the C$^{18}$O data in this box allows us to focus in on the high column density material 
lying in the immediate vicinity of the outflows. We measure $l_{2}$/$l_{1}$~=~0.18$\pm$~0.07
for the high column density material traced by C$^{18}$O, which is substantially smaller than
the global $l_{2}$/$l_{1}$ values found using $^{12}$CO and $^{13}$CO, but still reasonably
consistent with turbulence driven at large scales. Some caution should be applied to this result,
because, as noted above, large temporal variations in $l_{2}$/$l_{1}$ can occur in the case
of large scale driving. With this proviso, according to our model
results, the measured $l_{2}$/$l_{1}$ for the central region would imply a fractional 
driving scale of $\lambda_{D}/L_{c}$~$\approx$~0.5--1.0,
or a physical driving scale of 0.43--0.86~pc, based on the measured $l_{1}$~=~0.86~pc for the
C$^{18}$O data.  For reference, the cavity sizes of 0.1--0.2~pc in NGC~1333, if taken as a measure of the
driving scale within the 0.86~pc C$^{18}$O central core region, best match our models driven at 
$k_{d}$~=~3--4, for which we find $l_{2}$/$l_{1}$~$\approx$~0.11.
Examination of the C$^{18}$O second eigenimage structure reveals that it
shares, to some degree, the same north-south ``dipole'' structure seen in the $^{12}$CO and $^{13}$CO
second eigenimages. The presence of this signature, along with the $l_{2}$/$l_{1}$~=~0.18 measurement,
suggests that both the large-scale turbulence in the cloud as a whole, and small-scale (outflow) 
driven turbulence are important in this region. The inferred driving scale is therefore likely an
intermediate value between that arising from the outflows and that deriving from the large-scale
turbulent gradient across the core region. As a caveat, we note that the $^{12}$CO and $^{13}$CO 
lines are likely to better trace lower density, more spatially extended material than that traced 
by the C$^{18}$O line, so the relationship between the gradients seen in Figure~\ref{fig:eigim1333}
may not be as obvious as we assume. If the C$^{18}$O gradient is itself caused by outflow activity,
then this may indicate an interesting connection between the large and small-scale energy injection
mechanisms.

\begin{figure*}
\centering
\includegraphics[width=17cm]{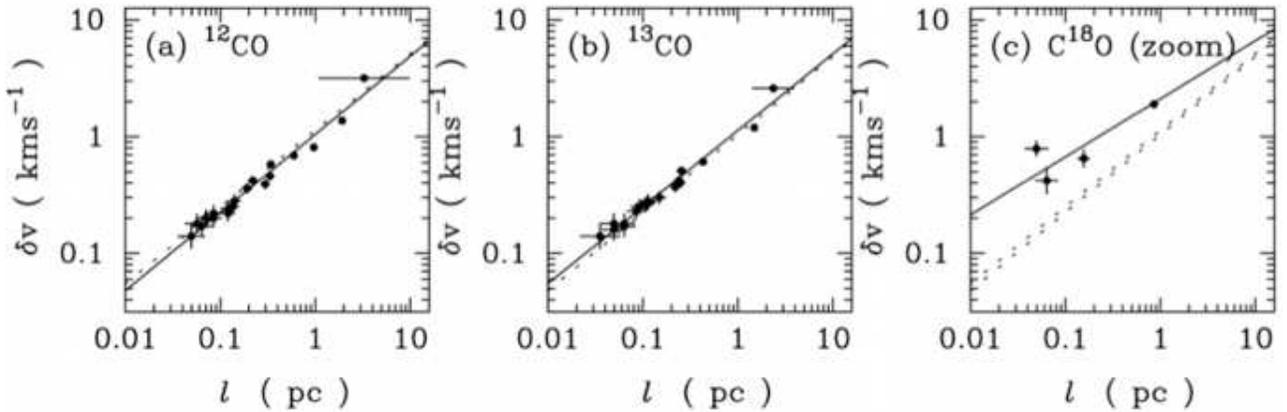}
   \caption{ Plots of $\delta{v}$ versus $l$ using all space and velocity scales obtained from principal component analysis of $^{12}$CO, $^{13}$CO, and C$^{18}$O J=1--0 data in the NGC~1333 moleclar cloud. The solid lines in each plot show the bisector fit to all points; the fitted relationships for $^{12}$CO and $^{13}$CO are repeated as dotted lines in all plots.  }
      \label{fig:dvl1333}
\end{figure*}

To examine the overall scale-dependence of turbulent motions in NGC~1333, in Figure~\ref{fig:dvl1333} 
we show plots of $\delta{v}$ versus $l$ for each isotope from their respective maps (see Brunt \& Heyer 2000b). 
It is noteworthy that the $^{12}$CO and $^{13}$CO measurements conform to the typical $\delta{v}$--$l$ relationship found
by Heyer \& Brunt (2004). Of more interest here is the increase in $\delta{v}$ seen on scales of $\sim$~0.1~pc
in the C$^{18}$O data, relative to the overall level set by the $^{12}$CO and $^{13}$CO data for the cloud as a
whole. This excess kinetic energy likely derives from the effect of outflows in the central core region
of the cloud. Although the number of retrieved $\delta{v}$--$l$ pairs is small, the data are in broad agreement
with Quillen et al (2005) who estimate outflow-driven cavity sizes and velocity perturbations of $\sim$~0.1--0.2~pc and
$\sim$~1~kms$^{-1}$ respectively. Thus internal driving of turbulence can be important in sub-parsec
regions of larger clouds, where a large number of outflows can develop. The PCA results for the NGC~1333 cloud 
as a whole set this in context, revealing the presence of larger-scale turbulence that will evolve on 
longer time scales than that present in the central core region. Turbulent dissipation in the dense part can
therefore be replenished not only by local sources, but by external ``driving'' by larger scale flows originating
in the surrounding cloud, as part of the overall hierarchy of turbulent motions. One cannot then consider 
the central star-forming regions as closed systems, evolving independently of their larger scale surroundings. 
Our cloud sample as a whole does not support a picture in which large scale turbulent motions have decayed 
sufficiently so that small scale driving alone is dominant. We conclude that either the clouds are continually 
driven on large scales, or that most clouds are sufficiently young that the initial seeding of turbulence 
by the large scale flows that created the cloud has not yet dissipated. Other observations (Ossenkopf \& Mac Low 2002;
Brunt \& Mac Low 2004) support this conclusion. It is not yet clear whether clouds are continually driven,
or whether the turbulence is in a decaying state. Offner, Klein, \& McKee (2008) find that while their simulated
clouds do not readily distinguish between decaying or driven conditions, there is a marginal preference for
continual driving. If clouds are driven at large scales, 
the turbulent dissipation time is comparable to their dynamical time (Mac Low 1999).

As noted above, the dipole pattern in the second eigenimage that is observed in all molecular clouds provides 
an important constraint to candidate driving sources. The dipole reflects the spatial distribution
of the largest velocity differences within a cloud. One can not directly
discriminate whether these velocity differences are due to shear, compressive, or
expanding motions. Large scale driving can readily account for such a pattern as it directly
deposits the energy at these scales. Turbulence driven on small scales can in principle
provide support on larger scales (Klessen {\it et al} 2000) and it may be possible for
excess small scale energy input to drive large scale expasion motion. More generally, a cluster-forming
clump could experience expansion, collapse, or perhaps oscillation about an equilibrium state,
depending on how active the star formation is. However, the dipole pattern suggests a more
directed flow of material, which would require the combined action of outflows to act in
a preferred direction. There is a possible mechanism for outflows to be oriented in a particular
direction: a strong magnetic field could result in core collapse along field lines, leading to co-oriented 
protostellar disks and therefore co-oriented outflows, for which some evidence is presented
in Anathpindika \& Whitworth (2008). The magnetic field strength needed to impose such directivity
is likely to inhibit cluster formation, and instead promote star formation in a more distributed,
quiescent mode (Heitsch, Mac Low, \& Klessen 2001; Price \& Bate 2008). Outflows from newborn stars and H{\sc ii} regions
can also redistribute energy from small to larger scales by driving expanding shells.  Such flows may also perturb
the magnetic field that threads the molecular cloud to excite Alfv\'en waves that can further
redistribute the outflow energy. However, such activity would again require implausible coherence
of location and alignment of outflows to reproduce the observed dipole pattern.

Another candidate for driving large scale turbulence ``internally'' is energy
injection by H{\sc ii} regions, as argued by Matzner (2002). However, large scale driving
is applicable to molecular clouds where H{\sc ii} regions are absent, such
as G216-2.5 (Maddelena's Cloud; Heyer, Williams, \& Brunt 2006). So
while these mechanisms are no doubt present in some molecular clouds, they cannot explain molecular
cloud turbulence in general and their effects will be limited to small scales.
If H{\sc ii} regions become large enough to drive large scale motions, then it is
likely that the cloud will be destroyed through photoionization rather than
``driven'' (Matzner 2002; Dale {\it et al}{} 2005).

Turbulence driven at large scales promotes star formation that is clustered, rapid, and
efficient, while small scale driving tends to form stars singly, slowly, and inefficiently
(Klessen {\it et al}{} 2000). If the star formation rate can be retarded by (additional) small scale
energy injection, it must do this in an environment which can be significantly (perhaps
dominantly) influenced by large scale turbulent flows of material. While the large scale versus
small scale driving picture can be modified by the effects of magnetic fields (Nakamura \& Li 2008; 
Price \& Bate 2008), it is in a much more dynamic way than that described by the quasistatic model 
(Shu, Adams, \& Lizano 1987). For example, recent high spatial dynamic range imaging of the Taurus molecular
cloud (Goldsmith {\it et al}{} 2008; Heyer {\it et al}{} 2008) reveal large scale,
magnetically-regulated, turbulent flows of material.

In addition to energy injection, another important consideration is the dissipation of turbulence.
Basu \& Murali (2001) argue that it is difficult to reconcile the inferred heating rate arising
from dissipation of turbulence with observed cloud luminosities unless the driving occurs at
large scales. More recently, Pan \& Padoan (2008) show that (assuming large scale driving)
heating by turbulent dissipation can exceed cosmic ray heating, and typical temperatures 
of $\sim$8.5~K can be sustained by turbulent heating alone. Since the turbulent heating
rate scales as $(\lambda_{D}/L_{c})^{-1}$,  widespread small-scale driving could lead to high cloud
temperatures that are incompatible with observations for molecular clouds as a whole (although
not for small sub-regions within the clouds). 

We mention a note of caution regarding the results presented here. The numerical
simulations of turbulence relied on random forcing (in Fourier space) to generate the
turbulent driving, which does not in detail adequately represent many physical sources of
energy injection. In the case of outflow-generated turbulence, considered here to be ``small
scale'', it was indeed found that the turbulence was effectively driven on small scales.
The close correspondence between the numerical models and the simple models of Paper I suggest 
also that it is not necessarily the details of the flows that are essential, but simply
the range of scales on which the turbulence is present. In this sense, the modeling completed
so far (Paper I and this work) adequately represent, statistically, turbulence with an
outer scale that is detectable in observations. It is to be expected that more
realistic driving mechanisms (e.g. as implemented by Nakamura \& Li 2007) can be
investigated in future. Finally, our results recommend that simulations of randomly forced
turbulence must necessarily include large scale driving in order to replicate real
molecular clouds. How this translates in detail to more realisitic driving
mechanisms must be addressed in future work.

\section{Summary}

We have examined simulated observations of the density and velocity fields from numerical simulations
of interstellar turbulence to investigate the scale at which energy is fed into molecular clouds. Using
principal component analysis, an observational measure of the driving scale can be obtained through the
ratio of characteristic scales of the second and first eigenimages. The measured ratio of eigenimage scales, 
$l_{2}/l_{1}$, has the same dependence on the normalized driving scale ($\lambda_{D}/L_{c}$) as derived for 
the normalized outer scale ($\lambda_{0}/L_{c}$) in the fBm models computed by Brunt (2003a).

Values of $l_{2}/l_{1}$ computed from spectroscopic imaging observations 
of molecular clouds are consistent with turbulence driven by large scale injection of energy. We have
examined a sample of 35 molecular clouds, and find that large-scale driving of turbulence provides
the best match for the sample as a whole. Detailed examination of the NGC~1333 cloud shows that 
this cloud as a whole is best described by large-scale driving, but that the central core regions
have been influenced by small-scale driving by outflows. However, while small scale driving of turbulence
through outflows can be important on small spatial scales on short time scales, it is not capable of 
reproducing the observed dipole structure of the second eigenimage.

The turbulence in our models was driven by random forcing, which will not 
represent energy injection by point-like sources very well, and future work on this issue should include
more realistic methods of driving turbulence. In the meantime, we recommend that turbulence simulations
that employ random forcing should ensure that the turbulence is driven on large scales to better 
recreate the dynamical conditions present in molecular clouds.

\begin{acknowledgements}
This work was supported by STFC Grant ST/F003277/1 to the University of Exeter,
Marie Curie Re-Integration Grant MIRG-46555 (CB), and NSF grant AST 0838222 to the Five College
Radio Astronomy Observatory. M-MML is supported by NSF CAREER grant 
AST99-85392 and NASA Astrophysical Theory Program grant NAG5-10103.  
Computations analyzed here were performed at the Rechenzentrum 
Garching of the Max-Planck-Gesellschaft. CB is supported by an RCUK fellowship
at the University of Exeter, UK.  We would like to thank Vesna Zivkov for assistance with the
simulated observations, Matthew Bate and Daniel Price for helpful discussions, and the anonymous 
referee for a number of interesting suggestions that improved the paper.
\end{acknowledgements}

\end{document}